\documentclass[twocolumn]{bmcart}

\usepackage[utf8]{inputenc} 

\usepackage{amsmath}
\usepackage{graphicx}
\usepackage{hyperref}

\usepackage{multicol}
\usepackage{multirow}
\usepackage{verbatim}

\usepackage{placeins} 
\usepackage[compatible]{algpseudocode} 
\usepackage[ruled,vlined,linesnumbered]{algorithm2e}

\usepackage[font={footnotesize},labelfont=bf]{caption}
\usepackage{threeparttable}
\usepackage{soul} 
\usepackage{xcolor} 

\startlocaldefs
\endlocaldefs

\begin{document}

\begin{frontmatter}

\begin{fmbox}
\dochead{Methodology Article}


\title{Joint Overlap Analysis of Multiple Genomic Interval Sets}


\author[
   addressref={aff1},                   
   corref={aff1},                       
   email={burcak@ceng.metu.edu.tr}   
]{\inits{BO}\fnm{Bur\c{c}ak} \snm{Otlu}}
 \author[
   addressref={aff1},
   email={tcan@ceng.metu.edu.tr}
]{\inits{TC}\fnm{Tolga} \snm{Can}}


\address[id=aff1]{
  \orgname{Department of Computer Engineering, Middle East Technical University}, 
  \postcode{06800}                                
  \city{Ankara},                              
  \cny{Turkey}                                    
}


\begin{artnotes}
\end{artnotes}



\begin{abstractbox}

\begin{abstract} 
\parttitle{Background} 
Next-generation sequencing (NGS) technologies have produced large volumes of genomic data. 
One common operation on heterogeneous genomic data is genomic interval intersection. 
Most of the existing tools impose restrictions such as not allowing nested intervals or requiring intervals to be sorted when finding overlaps in two or more interval sets.

\parttitle{Results}
We proposed segment tree (ST) and indexed segment tree forest (ISTF) based solutions for intersection of multiple genomic interval sets in parallel. 
We developed these methods as a tool, Joint Overlap Analysis (JOA), which takes $n$ interval sets and finds overlapping intervals with no constraints on the given intervals. 
The proposed indexed segment tree forest is a novel composite data structure, which
leverages on indexing and natural binning of a segment tree. 
We also presented construction and search algorithms for this novel data structure. 
We compared JOA ST and JOA ISTF with each other, and with other interval intersection tools for verification of its correctness and for showing that it attains comparable execution times.

\parttitle{Conclusions}
We implemented JOA in Java using the fork/join framework which speeds up parallel processing by taking advantage of all available processor cores. We compared JOA ST with JOA ISTF and showed that segment tree and indexed segment tree forest methods are comparable with each other in terms of execution time and memory usage. We also carried out execution time comparison analysis for JOA and other tools and demonstrated that JOA has comparable execution time and is able to further reduce its running time by using more processors per node.
JOA can be run using its GUI or as a command line tool. JOA is available with source code at \url{https://github.com/burcakotlu/JOA/}. 
A user manual is provided at \url{https://joa.readthedocs.org}
\end{abstract}



\end{abstractbox}
\end{fmbox}

\end{frontmatter}




\section{Background}\label{sec:background}
Genomic interval intersection is a major
component of analysis pipelines for Next Generation Sequencing
(NGS) technologies such as RNASeq, ChIPSeq, and exome sequencing.
There are various existing tools that perform interval
intersection (\cite{UCSCGenomeBrowser2016Update, BEDTools,
Pybedtools, BEDOPS, GROK}) and many other genomic analyses. UCSC Genome
Browser has been continuously improved since its first launch.
Lately, the Data Integrator feature is released in UCSC Genome
Browser, which allows users to combine and extract data from
multiple tracks (up to 5 tracks), simultaneously
\cite{UCSCGenomeBrowser2016Update}. BEDTools is a toolset developed for
genomics analysis tasks such as comparison, manipulation, and annotation of genomic features in
BAM, BED, GFF and VCF formats \cite{BEDTools}.
Pybedtools extends upon BEDTools and provides a Python
interface for BEDTools \cite{Pybedtools}.

BEDOPS is a highly scalable and easily-parallelizable genome
analysis toolkit, which enables tasks to be easily split by
chromosome for distributing whole-genome analyses across a
computational cluster \cite{BEDOPS}. BEDOPS requires sorted
interval set files before intersection. GROK utilizes region
algebra for genomic region operations and provides R, Python, LUA,
command line interfaces, and C++ API extension \cite{GROK}. NCList uses a novel data structure named Nested Containment List (NCList)
for interval databases \cite{NCList}. NCList keeps intervals in
lists such that intervals in each list have containment relation.
However, no list fully contains one another, which implies that if
lists are ordered on their start positions, then they are also
sorted on their end positions. GLANET stands for genomic loci annotation and enrichment tool and provides joint overlap analysis of 2 to 3 tracks at most \cite{GLANET}.

Layer {\em et al.} propose a novel parallel
``slice-then-sweep'' algorithm for $n$-way interval set
intersection. Their algorithm requires the input intervals not to
be contained fully in one another \cite{LayerParallelNWay}. 
Tabix indexes tab-delimited files
and converts a sequential access file into a random access file
\cite{Tabix}. Integrative Genomics Viewer
(IGV) is a high-performance visualization tool to explore genomic
datasets such as gene expression, ChIP-Seq, RNA-Seq, copy number
variations, and mutation data \cite{IGV}. Like Google Maps, it
allows users to zoom in and out on the whole genome up to single
base pair resolution.
		
BGT is a new, compact file format for efficiently storing and querying whole-genome genotypes of tens to hundreds of thousands of samples. 
BGT search alleles/variants in the input bgt format file within a given site or list of sites \cite{BGT}.
Genotype Query Tools (GQT) is a new indexing strategy that expedites analyses of genome variation datasets in VCF format based on sample genotypes, phenotypes and relationships \cite{GQT}. 
SeqArray is a storage-efficient high-performance new data format for WGS variant calls which comes with SeqArray software package including key R functions \cite{SeqArray}. 
BGEN is another novel binary file format for imputed genotype and haplotype data which are supported by many tools\cite{BGEN}.

In this paper, we generalize the problem of genomic interval
intersection (finding common overlapping intervals) from $2$ or
$3$ interval sets to $n$ interval sets. We implement segment tree
based construction and search algorithms in parallel for each
interval set and chromosome. We accept $n$ interval sets in
Browser Extensible Data (BED) format and we apply a divide and
conquer algorithm for finding common overlapping intervals. 
The intervals in a single set may overlap, may contain one another, and may be unsorted.
 We divide the interval sets into two sets, recursively, until one or two
interval sets remain as base cases. In case of two interval sets,
we construct a segment tree from one of the interval sets and we
then search the intervals of the other interval set as query intervals on the constructed segment tree.
In case of one interval set, we return all of its intervals
as overlapping intervals. Results coming from the base cases are
combined as follows: one of the results is used as query
intervals, a segment tree is built on the other result, query
intervals are searched on the segment tree, and overlapping
intervals are returned. This process is repeated recursively until
all of the $n$ interval sets are processed.

We also propose a composite data structure, indexed segment tree
forest which leverages on natural binning of segment trees and
indexing the segment tree nodes at a certain depth. The same
divide and conquer algorithm is also applied on the indexed
segment tree forest (ISTF). However, this time when a segment tree
is constructed, we proceed by converting this segment tree to an
indexed segment tree forest by cutting this tree at a certain
depth and indexing the segment tree nodes at that cut-off level.
Our aim is to reduce the search space and search on the shorter
indexed segment tree forest rather than one tall segment tree and
eventually to reduce the query time.
 The overview of the proposed methodology is depicted in Figure ~\ref{Figure1}.

\section{Methods}
\subsection{Segment Tree} Similar to interval trees, segment tree data structure is
another well-known space partitioning tree and a data structure
for storing intervals. Its space complexity is
$\mathcal{O}(nlog{n})$ and it can be constructed in
$\mathcal{O}(nlog{n})$ time for $n$ given intervals. Finding all
intervals in the segment tree containing a query point $q_{x}$
requires $\mathcal{O}(log{n}+k)$ time for $n$ given intervals and
$k$ hits  \cite{deBerg2010}.

Let $I := {[x_{1} : x'_{1}], [x_{2} : x'_{2} ], . . . , [x_{n} :
x'_{n}]}$ be a set of $n$ given intervals on the real line. Let
$p_{1}, p_{2}, . . . , p_{m}$ be the list of distinct interval
endpoints (low and high endpoints), sorted from left to right. We
simply partition the real line induced by these endpoints $p_{i}$. We
name the regions in this partitioning as elementary intervals.
Thus, the elementary intervals from these endpoints $p_{1}$, $p_{2}$,
..., $p_{m-1}$, $p_{m}$ are, from left to right, $[p_{1} : p_{1}],
(p_{1} : p_{2}), [p_{2} : p_{2}], . . . ,(p_{m-1} : p_{m}), [p_{m}
: p_{m}]$.


To this end, we build a binary search tree $T$ whose leaves
correspond to these elementary intervals induced by the endpoints
of the intervals in $I$ in an ordered way: the leftmost leaf
corresponds to the leftmost elementary interval, and so on. We
denote the elementary interval corresponding to a leaf $\mu$ by
$Int(\mu)$.

An internal node of $T$ represents the intervals that are the
union of intervals of its two children:  $Int(\nu)$ corresponding
to node $\nu$ is the union of the intervals $Int(\nu_{leftChild})$
and $Int(\nu_{rightChild})$ in the subtree rooted at $\nu$. The
interval represented by the parent of leaf nodes, $Int(\nu)$, is
the union of the elementary intervals $Int(\nu_{leftChild})$ and
$Int(\nu_{rightChild})$. Note that elementary intervals are
different from the input intervals, $I$.

Each internal node $\nu$ in $T$ represents an interval $Int(\nu)$
whereas each leaf node $\mu$ represents an elementary interval,
$Int(\mu)$. Each node stores a subset of input intervals, i.e.,
the canonical subset, $I(\nu)$, where $I(\nu)\subseteq I$. This
canonical subset of node $\nu$ stores the intervals $[x : x'] \in
I$ such that $Int(\nu)\subseteq [x : x']$ and $Int(parent(\nu))
\not\subseteq [x : x']$. This implies that 
if interval $[x : x']$ is in the canonical subset of node $\nu$, $I(\nu)$, then 
$Int(\nu)$ is completely contained in $[x : x']$ whereas $Int(parent(\nu))$
is not contained in $[x : x']$. The constructed balanced binary tree $T$
is called a segment tree. This way of construction ensures
non-overlapping, totally consecutive intervals for the nodes at
any depth, from left to right. In other words, it provides a
natural binning at any depth of the tree. In Figure
~\ref{Figure2}, we exemplify how we store $4$ intervals in the
segment tree leaves and internal nodes which are constructed from
the endpoints of these $4$ given intervals. Moreover, in Figure
~\ref{Figure2}, the arrows from the nodes point to their canonical subsets.

\subsubsection{Segment Tree Construction Complexity Analysis} To
construct a segment tree for $n$ given intervals, we proceed as
follows: We sort the endpoints of $n$ intervals in $O(n \log n)$
time and define elementary intervals at each end point and between
each consecutive endpoints.
We then construct a binary tree on these elementary intervals,
where each interval is the union of its left and right child's
intervals (or elementary intervals). This process continues until
the extent of all of the intervals are represented by the root
node. The construction can be done bottom-up in linear time. In
the last phase, $n$ given intervals are stored in the canonical
subset of the nodes, when a node's interval, $Int(\nu)$, is fully
contained in an input interval. As a result, an interval can be
attached to more than one node and the number of intervals
attached to nodes decreases as we go up in the tree; since, a
node's interval, $Int(\nu)$, becomes larger as we get closer to
the root node.

\subsection{Segment Tree Query} A query is processed starting at the root node. If
the query point $q_{x}$ overlaps with the node's interval,
$Int(\nu)$, the associated intervals stored at that node,
$I(\nu)$, are output and the query continues on the left or right
child of that node, visiting one node per level of the tree. The
time complexity of a segment tree query is $O(\log n + k)$ where
$n$ is the number of intervals and $k$ is the number of
overlapping intervals in the segment tree for the query point
$q_{x}$ \cite{deBerg2010}.

\subsection{Indexed Segment Tree Forest}
After analyzing constructed segment trees for real data sets, we
observed that nodes at the top of the segment tree (approximately
top two thirds of the segment tree) do not store any intervals or
store only a few intervals in their canonical subsets. In other words, input intervals are mostly stored in the bottom nodes of the
segment tree.


Keeping the whole segment tree with a significant number of nodes
without any or with a few intervals might be unnecessary. Furthermore,
passing through all these nodes for each query in order to find
overlapping intervals may increase the query time. Instead of
having one tall segment tree, we can cut the segment tree at a
certain depth close to the bottom of the tree and have as many
short segment trees rooted at segment tree nodes present at this cut-off
depth plus the segment tree nodes with no children above this
cut-off depth.

\subsection{Indexed Segment Tree Forest Parameters}
\subsubsection{Cut-off Depth}
We decide on the cut-off depth by considering the total number of
intervals stored in canonical subsets of nodes at the cut-off
depth and above the cut-off depth. 
We find a depth, $d$, such that the number of intervals stored from root node to the nodes at this depth is greater than $0.5\%$ of the total number of intervals stored in the segment tree. We then choose $d+1$ as the cut-off depth.
The closer the cut-off depth to the bottom of the tree, the more segment trees there will be in the forest. In other words, cut-off depth determines the number of short segment trees in the forest.

{\bf Moving intervals that were stored in the nodes above the
cut-off depth:} All the intervals attached to the nodes that are
above the cut-off depth must be distributed to the nodes at the
cut-off depth. Specifically, if an interval is attached to a node
above the cut-off depth, then this interval must be attached to
its offspring nodes at the cut-off depth. If there is no offspring
node at the cut-off depth then there can be two cases. Case 1: If
the node storing the interval has no offspring, then we directly
add this node to our segment tree forest. Case 2: If the node has
offspring/s above the cut-off depth, then we attach the interval
to its lowest offspring nodes and add these lowest offspring nodes
to our segment tree forest, with a higher priority given to the
node closer to the cut-off depth in order to keep the order
consistent between the intervals of the nodes. Note that we do this extra
work for a small number of nodes.

{\bf Linking segment tree nodes at the cut-off depth to each
other:} To ensure fast access between consecutive segment tree
nodes at the cut-off depth, we connect segment tree nodes to each
other through forward and backward pointers. We call these nodes linked nodes (See Figure
~\ref{Figure3}).

As mentioned earlier, we cut the segment tree at the cut-off depth
and keep the segment tree nodes at this cut-off depth and the
nodes with no children above the cut-off depth in an indexed
segment tree forest. Each segment tree node at this cut-off depth
is in fact a root of a segment tree in the forest and we compute its hash index
using a hash function for each segment tree node and we store
[index,segment tree node] tuples in a map.

\subsubsection{Cut-off Depth Decision, Percentage Parameter}
We make use of percentage parameter in cut-off depth decision.
This percentage parameter determines the number of intervals that needs to be moved w.r.t. the total number of intervals stored in the segment tree. As this percentage parameter increases, the number of intervals that we accept to move increases, and we cut the original segment tree closer to the leaves. We analyzed how this percentage parameter affects the cut-off depth and how the cut-off depth affects the construction and query time. For this purpose, we used an input of $2.3$ million intervals and varied the percentage parameter from $0.5$ to $10.0$. We observed that as we increase the percentage parameter, the number of intervals that need extra movement increases and the cut-off depth decreases, as expected (See Figure ~\ref{Figure4}). In this context, cut-off depth is defined as depth from the leaf level. However, construction and query time increases slightly, most probably because of more individual nodes are added to the indexed segment tree forest data structure (See Figure ~\ref{Figure5}). Based on these results, we decide on cut-off depth by considering movement of $0.5\%$ of the total intervals. As a result, we internally set the percentage parameter to $0.5\%$. 

\subsubsection{Hash Function, Preset Value}
\label{HashFunctionAndPresetValue} To index the nodes at the cut-off depth we use one universal hash
function as shown in Equation ~\ref{HashIndexEquation}. By using this hash function, we index
these short segment trees and we access each short segment tree in
$\mathcal{O}(1)$ time instead of $\mathcal{O}(cut-off)$ time. The preset value in the
hash function determines the number of different hash indexes that
one can have.

\begin{equation}
\label{HashIndexEquation} hashIndex =
(node.interval.low/presetValue)
\end{equation}

 Smaller preset values result in many hash indexes with less number of segment trees with
the same index, therefore, less number of collisions. But, smaller
preset values may result in sequential search of more than one
segment tree, which is definitely not preferred. 
Conversely, bigger preset values result in less hash indexes with more number of segment tree nodes with the same index, which implies more number of collisions.
To efficiently handle collisions,
we construct a binary search tree (BST) from the segment tree
nodes with the same index and store the root of this BST in the hash map.
Parent nodes of these linked nodes in the BST
constitute the artificial nodes as shown in Figure
~\ref{Figure6}.

 This collision handling strategy implies that each
segment tree may be reached in
$\mathcal{O}(height(BST))$ time instead of $\mathcal{O}(height(OriginalSegmentTree))$ time. 
As long as the height of BST formed from these segment trees with the same index is less
than the height of the original segment tree, search in indexed segment tree forest will
be still less than search in one tall segment tree.

 Theoretically, when all the nodes at the cut-off depth have the same hash index; the same segment tree will be constructed for them. Therefore, the height of hash BST will be always less than or equal to the height of the original segment tree. To exemplify this situation, we carried out an analysis by using inputs of $2.3$ million intervals. We observed that average height of the hash BST is less than the average height of the original segment tree for each chromosome. We also averaged over all chromosomes and showed that the average height of hash BSTs is $12.9$ and the average height of the original segment tree is $19.4$ for varying percentages from $0.5$ to $10$ and for the preset value of $1000000$ as shown in Table ~\ref{AverageHeightofHashBSTsVersusOriginalSegmentTree}.

\begin{table}[!htbp]
\begin{center}
\fontsize{10}{11}\selectfont
\scalebox{0.80}{
\begin{tabular}
{| c | r | r | r | r |} \hline
& \multicolumn{4}{c|}{Average Height}   \\  \cline{2-5}
Percentage & \multicolumn{2}{c|}{Hash BSTs}  & \multicolumn{2}{c|}{Original Segment Tree}    \\ \cline{1-5}
0.5 & \multicolumn{2}{c|}{12.9695} & \multicolumn{2}{c|}{\multirow{20}{*}{19.4}}  \\ \cline{1-3}
1.0 & \multicolumn{2}{c|}{12.9823} & \multicolumn{2}{c|}{} \\ \cline{1-3}
1.5 & \multicolumn{2}{c|}{12.9842} & \multicolumn{2}{c|}{} \\\cline{1-3}
2.0 & \multicolumn{2}{c|}{\multirow{2}{*}{12.9838}} & \multicolumn{2}{c|}{} \\ 
2.5 & \multicolumn{2}{c|}{} & \multicolumn{2}{c|}{} \\ \cline{1-3}
3.0 & \multicolumn{2}{c|}{12.9842} & \multicolumn{2}{c|}{} \\ \cline{1-3}
3.5 & \multicolumn{2}{c|}{\multirow{3}{*}{12.9845}} & \multicolumn{2}{c|}{} \\ 
4.0 & \multicolumn{2}{c|}{} & \multicolumn{2}{c|}{} \\ 
4.5 & \multicolumn{2}{c|}{} & \multicolumn{2}{c|}{} \\ \cline{1-3}
5.0 & \multicolumn{2}{c|}{\multirow{2}{*}{12.9843}} & \multicolumn{2}{c|}{} \\ 
5.5 & \multicolumn{2}{c|}{} & \multicolumn{2}{c|}{} \\\cline{1-3}
6.0 & \multicolumn{2}{c|}{12.9845} & \multicolumn{2}{c|}{} \\ \cline{1-3}
6.5 & \multicolumn{2}{c|}{12.9831} & \multicolumn{2}{c|}{} \\ \cline{1-3}
7.0 & \multicolumn{2}{c|}{\multirow{2}{*}{12.9821}} & \multicolumn{2}{c|}{} \\ 
7.5 & \multicolumn{2}{c|}{} & \multicolumn{2}{c|}{} \\ \cline{1-3}
8.0 & \multicolumn{2}{c|}{\multirow{5}{*}{12.9814}} & \multicolumn{2}{c|}{} \\ 
8.5 & \multicolumn{2}{c|}{} & \multicolumn{2}{c|}{} \\ 
9.0 & \multicolumn{2}{c|}{} & \multicolumn{2}{c|}{} \\ 
9.5 & \multicolumn{2}{c|}{} & \multicolumn{2}{c|}{} \\ 
10.0 & \multicolumn{2}{c|}{} & \multicolumn{2}{c|}{} \\ \cline{1-5}
\end{tabular}}
\end{center}
\caption{Average height of hash BSTs and original segment tree for all chromosomes are presented. While preset value of $1000000$ is kept constant, percentage parameter used in cut-off depth decision is varied from $0.5$ to $10.0$, and average height of hash BSTs and original segment tree are found using $2.3$ million intervals files. It is observed that average height of hash BSTs are always less than the average height of the original segment tree.} \label{AverageHeightofHashBSTsVersusOriginalSegmentTree}
\end{table}

\subsection{Query in Indexed Segment Tree Forest} For each query
interval, we compute its $lowIndex$ and $highIndex$ using its low
and high endpoints, respectively. We start searching on a linked
node pointed by the $lowIndex$ if it exists, otherwise we find the
$lowerIndex$ (highest index lower than $lowIndex$) and start
searching at the node shown by the $lowerIndex$ and continue
searching forward. If it is not possible, we start searching on
the linked node pointed by the $highIndex$ if it exists, if not, we
compute $higherIndex$ (lowest index higher than $highIndex$) and
search the node pointed by $higherIndex$ and continue searching
backward. If there is no node pointed by $higherIndex$, it means
that there is no overlapping interval with the query interval. 
The pseudocode of the indexed segment tree forest search algorithms are provided in the Supplementary Material.


\subsubsection{How to Guarantee that at Most Two Additional Index
Searches Are Enough?} As it is shown in
Figure~\ref{Figure7}, we first compute $lowIndex_{i}$
and $highIndex_{j}$ using query low and high endpoints,
respectively. 
Then we search for the segment trees pointed by one of these indexes in the order of  $lowIndex_{i}$, $lowIndex_{i-1}$, $highIndex_{j}$ or $highIndex_{j+1}$.

Here we present why we may need to consider only two more segment
trees pointed by the indexes $lowIndex_{i-1}$ and the
$highIndex_{j+1}$ (Figure ~\ref{Figure7}).

\FloatBarrier
\begin{equation}
lowIndex_{i}=queryLow/presetValue
\end{equation}

\begin{equation}
highIndex_{j}=queryHigh/presetValue
\end{equation}

\begin{equation}
lowIndex_{i-2} < lowIndex_{i-1} <  lowIndex_{i}
\end{equation}


\begin{equation}
lowIndex_{i-1} <  lowIndex_{i} \Rightarrow
\end{equation}

\begin{equation}
\label{equation1} lowNode_{i-1}.interval.low < queryLow
\end{equation}

From the preserved order between intervals of consecutive nodes we
know that

\begin{equation}
\begin{split}
    \label{equation2} lowNode_{i-2}.interval.high < \\
lowNode_{i-1}.interval.low
\end{split}
\end{equation}

Equations ~\ref{equation1} and ~\ref{equation2} imply that

\begin{equation}
lowNode_{i-2}.interval.high < queryLow \label{cannotoverlap}
\end{equation}

As a result of inequality ~\ref{cannotoverlap},
$lowNode_{i-2}.interval$ and query interval can not overlap.
Therefore we may need to look at only one more index preceding the
$lowIndex_{i}$ and search for the segment tree pointed by that
index and forward. In the same manner, we may need to consider
only one more index subsequent to the $highIndex_{j}$ and search
for the segment tree pointed by that index and backward.

\section{Results}
\subsection{Execution Time Comparison}
We have compared JOA with other tools using their latest available versions such as GROK v1.1.1, BEDTools v2.27.1, and BEDOPS  2.4.35. 
To compare the tools, we have used real and semi-synthetic datasets.
As an additional case study, we compared JOA segment tree and JOA indexed segment tree forest using 141 ENCODE Dnase hypersensitive sites.
Details of these datasets can be found in Availability of data and material. 

During comparisons, JOA is run both with segment tree (ST) and indexed segment tree forest (ISTF) options, GROK is run through its python API and intersectionL method is utilized, BEDTools is called by its  intersect -a -b utility, and for BEDOPS, --intersect set operation is used. 
BEDTools is run with its ``multiinter'' parameter when there are more than 2 input files.  
BEDTools gives a number to each input file starting at 1, when run with the ``multiinter'' parameter and produces a comma separated list of the input file numbers that are jointly overlapping in the fifth column of its output file.  
Therefore, from BEDTools' output file, we counted the number of rows that has all the file numbers separated by comma in the fifth column and calculated the number of jointly overlapping intervals.

All of the runs are carried out on Centos 6.6, Intel Xeon Gold 6132  (skylake), 2.6GHz $\times$ 28 processors (14 cores/socket).
All the provided execution times in Table ~\ref{RealData_Benchmarks_AverageRuntimes_Table}, ~\ref{Simulations_Scenario1_AverageRuntimes_Table}, ~\ref{Simulations_Scenario2_AverageRuntimes_Table} and ~\ref{AdditionalCaseStudyforJOATable} are averaged over 50 runs.

\subsubsection{Execution Time Comparison of Tools Using Benchmarks Datasets} 
The first two benchmark datasets include five files that are hotspot peaks for five fetal adrenal tissues: fAdrenal-DS12528, fAdrenal-DS15123, fAdrenal-DS17319, fAdrenal-DS17677 and fAdrenal-DS20343. They contain 193835, 188966, 137386, 132500, and 195098 intervals, respectively. The benchmark that is referred to as ``2 small files'' are the first two files of these five files.

 The rest of the benchmark datasets include $2,362,386$ (2.3 M), $6,473,749$ (6.4 M), and $9,218,913$ (9.2 M) intervals, respectively. Input of the last benchmark is the last 2 BED files with 6.4 M and 9.2 M intervals, respectively.

\begin{table*}[!htbp]
\begin{center}
\fontsize{10}{11}\selectfont
\scalebox{0.80}{
\begin{tabular}
{| c | r | r | r | r | r | r | r |} \hline
\multicolumn{1}{|c|}{\multirow{4}{*}{\parbox{3.5cm}{\centering Benchmarks}}}& \multicolumn{7}{c|}{Average execution times in seconds}    \\  \cline{2-8}
\multicolumn{1}{|c|}{} & \multicolumn{5}{c|}{(1n,1ppn)} & \multicolumn{2}{c|}{(1n,8ppn)}   \\  \cline{2-8}
\multicolumn{1}{|c|}{} & \multicolumn{1}{c|}{\multirow{2}{*}{GROK}} &  \multicolumn{1}{c|}{\multirow{2}{*}{BEDOPS}} & \multicolumn{1}{c|}{\multirow{2}{*}{BEDTools}} & \multicolumn{1}{c|}{\multirow{2}{*}{JOA ST}} & \multicolumn{1}{c|}{\multirow{2}{*}{JOA ISTF}} & \multicolumn{1}{c|}{\multirow{2}{*}{JOA ST}} & \multicolumn{1}{c|}{\multirow{2}{*}{JOA ISTF}}   \\
\multicolumn{1}{|c|}{} & \multicolumn{1}{c|}{} & \multicolumn{1}{c|}{} & \multicolumn{1}{c|}{} & \multicolumn{1}{c|}{} &  \multicolumn{1}{c|}{} & \multicolumn{1}{c|}{} & \multicolumn{1}{c|}{} \\ \cline{1-8}
2 small files 	& 1.1426 	& 0.1785		& 0.4204					& 1.5135 	& 1.7880 	& 0.8363 	& 0.9712 	\\ \cline{1-8}
5 small files 	& 2.3757 	& 0.3104		& 12.6509				& 2.1504 	& 2.7396 	& 1.2470 	& 1.5531 	\\ \cline{1-8}
2.3 M 			& 13.5173  	& 1.8656		& 14.4679				& 16.1611 	& 17.0239 	& 9.5708 	& 9.7355 	\\ \cline{1-8}
6.4 M 			& 34.2736 	& 4.8053		& 63.8499				& 39.7799 	& 43.5383 	& 23.9280 	& 24.0116 	\\ \cline{1-8}
9.2 M 			& 54.5452 	& 7.1102		& 65.3324				& 56.0001 	& 55.9838 	& 31.7290 	& 32.5613 	\\ \cline{1-8}
(6.4,9.2) M 		& 49.2007 	& 5.4660		& 43.1104				& 48.0046 	& 47.9815 	& 23.2048 	& 24.3927 	\\ \cline{1-8}
\end{tabular}}
\end{center}
\caption{Average execution time comparison of the tools for the six benchmarks. GROK, JOA, BEDTools and BEDOPS execution times are averaged over 50 runs.} \label{RealData_Benchmarks_AverageRuntimes_Table}
\end{table*}

JOA ST and ISTF are run for both (1 node, 1 processor per node) and (1 node, 8 processors per node) settings.  
Average execution times are listed in Table ~\ref{RealData_Benchmarks_AverageRuntimes_Table}. 
Running times are comparable and usage of more processors per node decreased the running time of JOA because of parallel implementation of ST and ISTF algorithms.
For ``5 small files'', BEDTools ``multiinter'' parameter usage increased its running time as it has a different implementation which is also reflected in its output.
Number of overlapping intervals found for each benchmark dataset is listed in Table ~\ref{RealData_Benchmarks_NumberOfOverlappingIntervalsFound_Table}. 
JOA and BEDTools found exactly the same number of overlapping intervals whereas GROK and BEDOPS found different number of overlapping intervals. This might be because of a different design, implementation or a case which is handled differently such as nested intervals.

\begin{table*}[!htbp]
\begin{center}
\fontsize{10}{11}\selectfont
\scalebox{0.8}{
\begin{tabular}
{| c | r | r | r | r | r |} \hline
\multicolumn{1}{|c|}{\multirow{3}{*}{Benchmarks}}& \multicolumn{5}{c|}{Number of overlapping intervals found}   \\  \cline{2-6}
\multicolumn{1}{|c|}{} & \multicolumn{1}{c|}{\multirow{2}{*}{GROK}} &  \multicolumn{1}{c|}{\multirow{2}{*}{BEDOPS}} & \multicolumn{1}{c|}{\multirow{2}{*}{BEDTools}} & \multicolumn{1}{c|}{\multirow{2}{*}{JOA ST}} & \multicolumn{1}{c|}{\multirow{2}{*}{JOA ISTF}}   \\ 
\multicolumn{1}{|c|}{} & \multicolumn{1}{c|}{} & \multicolumn{1}{c|}{} & \multicolumn{1}{c|}{} & \multicolumn{1}{c|}{} &  \multicolumn{1}{c|}{} \\ \cline{1-6}
2 small files 	& 145925  	& 145925  	& 145925  	& 145925 	& 145925   \\ \cline{1-6}
5 small files 	& 93080 		& 93080 		& 93080 		& 93080  	& 93080    \\ \cline{1-6}
(2.3 M, 2.3 M) 	& 2972223 	& 753639  	& 13903684 	& 13903684 	& 13903684 \\ \cline{1-6}
(6.4 M, 6.4 M) 	& 7369608  	& 1292747  	& 47936947 	& 47936947 	& 47936947 \\ \cline{1-6}
(9.2 M, 9.2 M) 	& 11553395 	& 3046101 	& 39217773 	& 39217773  	& 39217773 \\ \cline{1-6}
(6.4 M, 9.2 M) 	& 4746709 	& 424074 	& 19645676 	& 19645676  	& 19645676 \\ \cline{1-6}
\end{tabular}} 
\end{center}
\caption{Number of overlapping intervals found for each benchmark. JOA and BEDTools found exactly the same number of overlapping intervals whereas GROK and BEDOPS found different number of overlapping intervals.}
\label{RealData_Benchmarks_NumberOfOverlappingIntervalsFound_Table}
\end{table*}

\subsubsection{Execution Time Comparison of Tools Using Semi-Synthetic Datasets}
To compare the tools on a controlled dataset, we sampled intervals of each 500 bp long from the human genome uniformly at random and generated bed format input.
In the first scenario, we kept the number of intervals constant at $100,000$ in each file and we increased the number of files by twofold from $2$ to $512$.
In the second scenario, we kept the number of files constant at $2$ and we increased the number of intervals in each file by twofold from $1,000,000$ to $16,000,000$.

In the first scenario, we could not run GROK for more than 256 input files since calling a python function with more than 256 arguments was not allowed.
Also, in the first scenario, BEDTools is called with the ``multiinter'' parameter when there are more than two input files.

Average execution times for both scenarios are provided in Tables ~\ref{Simulations_Scenario1_AverageRuntimes_Table} and ~\ref{Simulations_Scenario2_AverageRuntimes_Table}.
In the first scenario, when the number of input files is greater than 2, BEDTools average execution times are extremely high because of the ``multiinter'' option usage. 
In both scenarios, JOA has reduced execution times when 2 processors per node (ppn) is used instead of 1 ppn.

Number of overlapping intervals found for the first scenario and the second scenario are provided in Tables ~\ref{Simulations_Scenario1_NumberofOverlappingIntervalsFound_Table} and ~\ref{Simulations_Scenario2_NumberofOverlappingIntervalsFound_Table}, respectively.
JOA and BEDTools have found the same number of overlapping intervals for the semi-synthetic datasets as in the case of benchmark datasets.
However, GROK and BEDOPS have found different number of overlapping intervals.

\begin{table*}[!htbp]
\begin{center}
\fontsize{10}{11}\selectfont
\scalebox{0.80}{
\begin{tabular}
{| c | r | r | r | r | r | r | r |} \hline
\multicolumn{1}{|c|}{\multirow{4}{*}{\parbox{3.5cm}{\centering Simulations\\$1^{st}$ Scenario\\(\#ofFiles,\#ofIntervals)}}}& \multicolumn{7}{c|}{Average execution times in seconds}    \\  \cline{2-8}
\multicolumn{1}{|c|}{} & \multicolumn{5}{c|}{(1n,1ppn)} & \multicolumn{2}{c|}{(1n,2ppn)}   \\  \cline{2-8}
\multicolumn{1}{|c|}{} & \multicolumn{1}{c|}{\multirow{2}{*}{GROK}} &  \multicolumn{1}{c|}{\multirow{2}{*}{BEDOPS}} & \multicolumn{1}{c|}{\multirow{2}{*}{BEDTools}} & \multicolumn{1}{c|}{\multirow{2}{*}{JOA ST}} & \multicolumn{1}{c|}{\multirow{2}{*}{JOA ISTF}} & \multicolumn{1}{c|}{\multirow{2}{*}{JOA ST}} & \multicolumn{1}{c|}{\multirow{2}{*}{JOA ISTF}}   \\ 
\multicolumn{1}{|c|}{} & \multicolumn{1}{c|}{} & \multicolumn{1}{c|}{} & \multicolumn{1}{c|}{} & \multicolumn{1}{c|}{} &  \multicolumn{1}{c|}{} & \multicolumn{1}{c|}{} & \multicolumn{1}{c|}{} \\ \cline{1-8}
(2,100000) 	& 0.6973 	& 0.0366		& 0.2380			& 1.0846 	& 1.2861		& 0.6677 	& 0.7121\\ \cline{1-8}
(4,100000) 	& 1.3527 	& 0.0397		& 3.8666 		& 1.4480 	& 1.7342 	& 1.0410 	& 1.0932\\ \cline{1-8}
(8,100000)	& 2.7232  	& 0.0447		& 8.2620 		& 2.0175  	& 2.4720  	& 1.4686 	& 1.5197\\ \cline{1-8}
(16,100000) 	& 5.7325 	& 0.0565		& 18.9487 		& 2.8402 	& 3.6387  	& 2.0252 	& 2.3860\\ \cline{1-8}
(32,100000) 	& 12.5799  	& 0.0819		& 45.7564 		& 4.4840 	& 5.4132  	& 3.0647 	& 3.6630\\ \cline{1-8}
(64,100000) 	& 30.2584 	& 0.1283		& 116.7736 		& 6.9214 	& 7.9966 	& 4.5746  	& 5.4844\\ \cline{1-8}
(128,100000) & 66.7122 	& 0.2131		& 312.8438	 	& 12.1682  	& 13.6740 	& 7.4285 	& 8.3240\\ \cline{1-8}
(256,100000) & \multicolumn{1}{c|}{-} 		& 0.4229		& 891.2454  	& 22.6010 	& 24.0628  	& 15.3915 	& 16.2595\\ \cline{1-8}
(512,100000) & \multicolumn{1}{c|}{-} 		& 0.8283		& 2887.3816 	& 42.0187  	& 45.3321 	& 26.9868 	& 28.8161\\ \cline{1-8}
\end{tabular}}
\end{center}
\caption{Number of interval files are increased by two fold whereas number of intervals in each file is kept constant. Average execution time comparison of the tools for the first scenario using semi-synthetic datasets. GROK, JOA, BEDTools and BEDOPS execution times are
averaged over 50 runs except BEDTools runs for 128, 256 and 512 files which are averaged over 2 runs.} \label{Simulations_Scenario1_AverageRuntimes_Table}
\end{table*}

\begin{table*}[!htbp]
\begin{center}
\fontsize{10}{11}\selectfont
\scalebox{0.80}{
\begin{tabular}
{| c | r | r | r | r | r |} \hline
\multicolumn{1}{|c|}{\multirow{3}{*}{\parbox{3.5cm}{\centering Simulations\\$1^{st}$ Scenario\\(\#ofFiles,\#ofIntervals)}}}& \multicolumn{5}{c|}{Number of Overlapping Intervals Found}    \\  \cline{2-6}
\multicolumn{1}{|c|}{} & \multicolumn{1}{c|}{\multirow{2}{*}{GROK}} &  \multicolumn{1}{c|}{\multirow{2}{*}{BEDOPS}} & \multicolumn{1}{c|}{\multirow{2}{*}{BEDTools}} & \multicolumn{1}{c|}{\multirow{2}{*}{JOA ST}} & \multicolumn{1}{c|}{\multirow{2}{*}{JOA ISTF}}   \\ 
\multicolumn{1}{|c|}{} & \multicolumn{1}{c|}{} & \multicolumn{1}{c|}{} & \multicolumn{1}{c|}{} & \multicolumn{1}{c|}{} &  \multicolumn{1}{c|}{}  \\ \cline{1-6}
(2,100000)	& 3958 	& 3761		& 3913		& 3913 	& 3913	\\ \cline{1-6}
(4,100000) 	& 5 		& 5			& 5 			& 5 	& 5 	\\ \cline{1-6}
(8,100000)	& 0  	& 0			& 0 			& 0 	& 0 \\ \cline{1-6}
(16,100000) & 0 		& 0			& 0 			& 0 	& 0 	\\ \cline{1-6}
(32,100000) & 0 		& 0			& 0 & 0 	& 0 	\\ \cline{1-6}
(64,100000) & 0 		& 0			& 0 & 0 	& 0 	\\ \cline{1-6}
(128,100000) & 0 	& 0			& 0 & 0 	& 0 	\\ \cline{1-6}
(256,100000) & \multicolumn{1}{c|}{-} 	& 0	& 0 & 0 	& 0 	\\ \cline{1-6}
(512,100000) & \multicolumn{1}{c|}{-} 	& 0	& 0 & 0 	& 0 	\\ \cline{1-6}
\end{tabular}}
\end{center}
\caption{Number of overlapping intervals found by the tools for the first scenario of semi-synthetic datasets.} \label{Simulations_Scenario1_NumberofOverlappingIntervalsFound_Table}
\end{table*}

\begin{table*}[!htbp]
\begin{center}
\fontsize{10}{11}\selectfont
\scalebox{0.80}{
\begin{tabular}
{| c | r | r | r | r | r | r | r |} \hline
\multicolumn{1}{|c|}{\multirow{4}{*}{\parbox{3.5cm}{\centering Simulations\\$2^{nd}$ Scenario\\(\#ofFiles,\#ofIntervals)}}}& \multicolumn{7}{c|}{Average execution times in seconds}    \\  \cline{2-8}
\multicolumn{1}{|c|}{} & \multicolumn{5}{c|}{(1n,1ppn)} & \multicolumn{2}{c|}{(1n,2ppn)}   \\  \cline{2-8}
\multicolumn{1}{|c|}{} & \multicolumn{1}{c|}{\multirow{2}{*}{GROK}} &  \multicolumn{1}{c|}{\multirow{2}{*}{BEDOPS}} & \multicolumn{1}{c|}{\multirow{2}{*}{BEDTools}} & \multicolumn{1}{c|}{\multirow{2}{*}{JOA ST}} & \multicolumn{1}{c|}{\multirow{2}{*}{JOA ISTF}} & \multicolumn{1}{c|}{\multirow{2}{*}{JOA ST}} & \multicolumn{1}{c|}{\multirow{2}{*}{JOA ISTF}}   \\ 
\multicolumn{1}{|c|}{} & \multicolumn{1}{c|}{} & \multicolumn{1}{c|}{} & \multicolumn{1}{c|}{} & \multicolumn{1}{c|}{} &  \multicolumn{1}{c|}{} & \multicolumn{1}{c|}{} & \multicolumn{1}{c|}{} \\ \cline{1-8}
(2,1M) 	& 9.0330 	& 0.3635		& 4.2109 	& 4.5382   	& 5.2108 	& 3.4880  	& 4.1231 \\ \cline{1-8}
(2,2M) 	& 19.7827 	& 0.6202		& 11.4695	& 11.8475  	& 12.2080  	& 7.2478  	& 8.2213 \\ \cline{1-8}
(2,4M)	& 44.9517   	& 1.8149		& 33.1372 	& 29.9437  	& 32.5128  	& 18.4192  	& 19.3616\\ \cline{1-8}
(2,8M) 	& 102.6299  	& 2.9645 	& 108.0360 	& 73.2382  	& 74.9382  	& 55.1262  	& 56.9878\\ \cline{1-8}
(2,16M) 	& 255.6354  	& 5.7609 	& 397.5899	& 190.2903  & 192.3511  	& 178.7225  & 185.8660\\ \cline{1-8}
\end{tabular}}
\end{center}
\caption{Number of intervals in each file is increased by two fold whereas number of interval files is kept constant. Average execution time comparison of the tools for the second scenario of semi-synthetic datasets. GROK, JOA, BEDTools and BEDOPS execution times are
averaged over 50 runs.} \label{Simulations_Scenario2_AverageRuntimes_Table}
\end{table*}

\begin{table*}[!htbp]
\begin{center}
\fontsize{10}{11}\selectfont
\scalebox{0.80}{
\begin{tabular}
{| c | r | r | r | r | r |} \hline
\multicolumn{1}{|c|}{\multirow{3}{*}{\parbox{3.5cm}{\centering Simulations\\$2^{nd}$ Scenario\\(\#ofFiles,\#ofIntervals)}}}& \multicolumn{5}{c|}{Number of Overlapping Intervals Found}   \\  \cline{2-6}
\multicolumn{1}{|c|}{} & \multicolumn{1}{c|}{\multirow{2}{*}{GROK}} &  \multicolumn{1}{c|}{\multirow{2}{*}{BEDOPS}} & \multicolumn{1}{c|}{\multirow{2}{*}{BEDTools}} & \multicolumn{1}{c|}{\multirow{2}{*}{JOA ST}} & \multicolumn{1}{c|}{\multirow{2}{*}{JOA ISTF}}   \\ 
\multicolumn{1}{|c|}{} & \multicolumn{1}{c|}{} & \multicolumn{1}{c|}{} & \multicolumn{1}{c|}{} & \multicolumn{1}{c|}{} &  \multicolumn{1}{c|}{}  \\ \cline{1-6}
(2,1M) 	& 432665 	& 280649	 & 402931	& 402931 	& 402931		\\ \cline{1-6}
(2,2M) 	& 1741459 	& 804859  & 1613107	& 1613107	& 1613107 	\\ \cline{1-6}
(2,4M)	& 6509486  	& 1785275 & 6455690	& 6455690 	& 6455690 	\\ \cline{1-6}
(2,8M) 	& 20936796 	& 2706065 & 25812190 	& 25812190 	& 25812190 	\\ \cline{1-6}
(2,16M) 	& 55354698 	& 2336735 & 103245490 & 103245490 	& 103245490 	\\ \cline{1-6}
\end{tabular}}
\end{center}
\caption{Number of overlapping intervals found by the tools for the second scenario of semi-synthetic datasets.} \label{Simulations_Scenario2_NumberofOverlappingIntervalsFound_Table}
\end{table*}

\subsubsection{Execution Time Comparison of JOA ST versus ISTF Using ENCODE data}
To show the parallel processing advantage of JOA, we designed and ran an additional case study for JOA using 141 interval sets of Dnase hypersensitive sites. 
We found and supplied all jointly overlapping intervals for these 141 interval sets in Supplementary Table S1. 
This additional case study verified that JOA can handle a significantly larger number of interval files and JOA is able to attain better running times with the help of its parallel processing ability as shown in Table ~\ref{AdditionalCaseStudyforJOATable}.

\begin{table}[!htbp]
\begin{center}
\fontsize{10}{11}\selectfont
\scalebox{0.8}{
\begin{tabular}
{| c | r | r | r | r |} \cline{1-5}

\multicolumn{1}{|c|}{\multirow{2}{*}{\parbox{3cm}{\centering ENCODE\\ 141 DHSs files}}} & \multicolumn{4}{|c|}{Average execution times in seconds}   \\  \cline{2-5}
\multicolumn{1}{|c|}{} & \multicolumn{2}{c|}{(1n,1ppn)} &  \multicolumn{2}{c|}{(1n,8ppn)}  \\  \cline{1-5}

JOA ST 		& \multicolumn{2}{c|}{15.8894} 	& \multicolumn{2}{c|}{9.6301}   	\\ \cline{1-5}
JOA ISTF 	& \multicolumn{2}{c|}{19.0009}	& \multicolumn{2}{c|}{10.6943}	\\ \cline{1-5}
\end{tabular}}
\end{center}
\caption{Average execution time comparison of JOA ST and ISTF for 141 interval sets of Dnase hypersensitive sites. JOA ST and ISTF execution times are averaged over 50 runs. Average execution times are reduced as number of processors per node is increased from 1 to 8.}
\label{AdditionalCaseStudyforJOATable}
\end{table}

\subsection{JOA Segment Tree versus Indexed Segment Tree Forest Detailed Execution Time Comparison}
 We used the interval sets of $2.3$ million, $6.4$ million, $9.2$ million, and $6.4\&9.2$ million intervals from benchmarks datasets. 
We compared JOA ST versus JOA ISTF with respect to read, construction and query running times which are averaged over 50 runs. 
We show that ST and ISTF perform comparably with respect o read, construction and query running times as it is shown in Figure ~\ref{Figure8}.

\subsection{JOA Memory Usage}
Segment tree requires $\mathcal{O}(nlog{n})$ storage for $n$ given intervals.
Since indexed segment tree forest is based on indexing of segment tree nodes at a certain depth, it has similar memory requirements. 
JOA is implemented in Java with fork/join framework which speeds up parallel processing by using all available processor cores. 
However, parallel processing requires loading all the data into memory.
Also, segment tree and indexed segment tree forest construction requires memory allocation of many temporary data structures which can be only deallocated automatically by the Java garbage collector.  

We calculated the memory usage of JOA ST and ISTF for the semi-synthetic datasets.
The memory usage in megabytes (MBs) are presented in Supplementary Tables S2 and S3. 
All the memory usage in the first and second scenarios are as expected; however, for the 2 interval sets of 8M intervals each, and for the 2 interval sets of 16M intervals in the second scenario, JOA's memory consumption is high. This can be the due to the loading all interval sets into memory, the additional temporarily used data structures, or the Java garbage collector. 
As long as Java Virtual Machine has enough memory, garbage collector may not take action to deallocate memory and follow manually written memory deallocation statements in the code.

\section{Discussion}
For both real and semi-synthetic datasets used in execution time comparisons, JOA ST, JOA ISTF and BEDTOOLs found same number of overlapping intervals.
However, GROK and BEDOPS found different number of overlapping intervals. 
This can be because of different implementation for the case of nested intervals.

An advantage of JOA ST and JOA ISTF is their parallel implementation. Nonetheless, the usage of 8 ppn instead of 1 ppn did not reduce the running time as expected but reduced it by half only.
This can be related to the intercommunication bottleneck that we can face as we increase the number of processors.

Parallel implementation of JOA ST and JOA ISTF algorithms require all the input intervals to be loaded into the memory.
Moreover, JAVA garbage collector deallocates memory automatically without taking manually written memory deallocations into account.
Therefore, although JOA ST and JOA ISTF have comparable and sometimes lower execution times, their memory footprint may be high.
 

 Regarding JOA ISTF, we analyzed how the percentage parameter affects the cut-off depth and how the cut-off depth affects the execution time (Figures ~\ref{Figure4} and ~\ref{Figure5}). Depending on these analyses, we set the percentage parameter to $0.5\%$. In addition to that, we provided a detailed analysis on the average height of the binary search trees (BSTs) constructed from hash indexes. We showed that average height of BSTs is always less than or equal the height of the original segment tree.

 Furthermore, for JOA ISTF, we analyzed the number of hash indexes for varying preset values
using human genome, chromosome 1 intervals of the 5 small files'  which is provided as the second benchmark dataset in Section 3.1.1. The smaller the preset value, the higher the number of different hash indexes, and vice versa (Figure ~\ref{Figure9}).

Moreover, we computed the mean and standard deviation of the number of
segment tree nodes assigned to the same hash index as we vary the preset
value. The smaller the preset value, the less the number of segment
tree nodes assigned to a hash index (Figure ~\ref{Figure10}).
This shows that there is a trade off between number of hash
indexes and the mean number of segment tree nodes assigned to a hash index as we change the preset value.
This is an inherent advantage of the proposed indexed segment tree forest construction algorithm and allows for generalizing to varying preset values.


\section{Conclusion}
 In this paper, we presented efficient methods for parallel joint overlap analysis of $n$ interval sets. The proposed segment tree and novel indexed segment tree forest solutions are optimized with a divide and conquer algorithm design and implemented as a tool named JOA. We showed that JOA ST and ISTF have comparable execution times. We compared JOA with other state of the art tools such as GROK, BEDTools, BEDOPS and demonstrated the efficacy of JOA with respect to the execution time. Especially, when the number of processors per node is increased, JOA had less execution time than the other tools. Furthermore, we also verified that JOA is able to identify all the overlapping intervals.

 To verify the parallel processing advantage of JOA, we designed and ran an additional case study for JOA in which we found the jointly overlapping intervals of 141 interval sets of ENCODE Dnase hypersensitive sites.



\begin{backmatter}

\section*{Declarations}
\subsection*{Abbreviations}
NGS: Next Generation Sequencing; JOA: Joint Overlap Analysis; ST: Segment Tree; ISTF: Indexed Segment Tree Forest

\subsection*{Ethics approval and consent to participate}
Not applicable.

\subsection*{Consent for publication}
Not applicable.

\subsection*{Availability of data and material}\label{AvailabilityOfDataAndMaterial}
JOA's source code is available at \url{https://github.com/burcakotlu/JOA/}. 
You can download JOA executable jar, joa.jar from \url{https://www.dropbox.com/s/vflr45vk2lnxk6h/joa.jar?dl=0}.\\
A user manual is provided at \url{https://joa.readthedocs.org}.
\newline \newline
Benchmarks Dataset:
The first two of benchmark datasets used in execution time comparison are hotspot peaks for five fetal adrenal issues: 
fAdrenal-DS12528, fAdrenal-DS15123, fAdrenal-DS17319, fAdrenal-DS17677 and fAdrenal-DS20343.
They are all in bed format and available for download from \url{http://burcak.ceng.metu.edu.tr/joa/}.
\\
Rest of the benchmark datasets, 2.3 M Intervals (wgEncodeBroadHistoneHepg2CtcfStdAlnRep1), 6.4 M Intervals (wgEncodeBroadHistoneDnd41H3k04me2AlnRep2), and 9.2 M Intervals (wgEncodeBroadHistoneA549H3k04me2Dex100nmAlnRep2) are downloaded as bam format files from \url{http://hgdownload.cse.ucsc.edu/goldenpath/hg19/encodeDCC/wgEncodeBroadHistone/} and converted into bed format files.
\newline \newline
	
Semi-Synthetic Dataset:
All files are in bed format. Compressed simulated\_data.tar.gz file can be downloaded from \url{https://drive.google.com/open?id=1Y7lWzBYEoDjz4l8wGDXkd-0m6-SBNVrM}. \newline \newline
ENCODE Data:
For additional case study of JOA, we downloaded 141 interval sets of ENCODE Dnase hypersensitive sites (29 of them are in pk format and 112 of them are in bed format) from \url{http://hgdownload.cse.ucsc.edu/goldenpath/hg19/encodeDCC/wgEncodeUwDnase/} and \url{http://hgdownload.cse.ucsc.edu/goldenpath/hg19/encodeDCC/wgEncodeOpenChromDnase/}.




\subsection*{Competing interests}
The authors declare that they have no competing interests.

\subsection*{Funding}
No funding was received for the study.

\subsection*{Author's contributions}
BO conceived the idea of indexed segment tree forest, designed and implemented the parallel algorithms for segment tree and indexed segment tree forest, conducted execution time comparison analysis and wrote the draft. TC supervised the project, execution time comparison analysis and revised the final draft.

\subsection*{Acknowledgements}
Not applicable.


\bibliographystyle{bmc-mathphys} 




\FloatBarrier
\section*{Figures}
\begin{figure}
\centering
\includegraphics[width=0.5\textwidth]{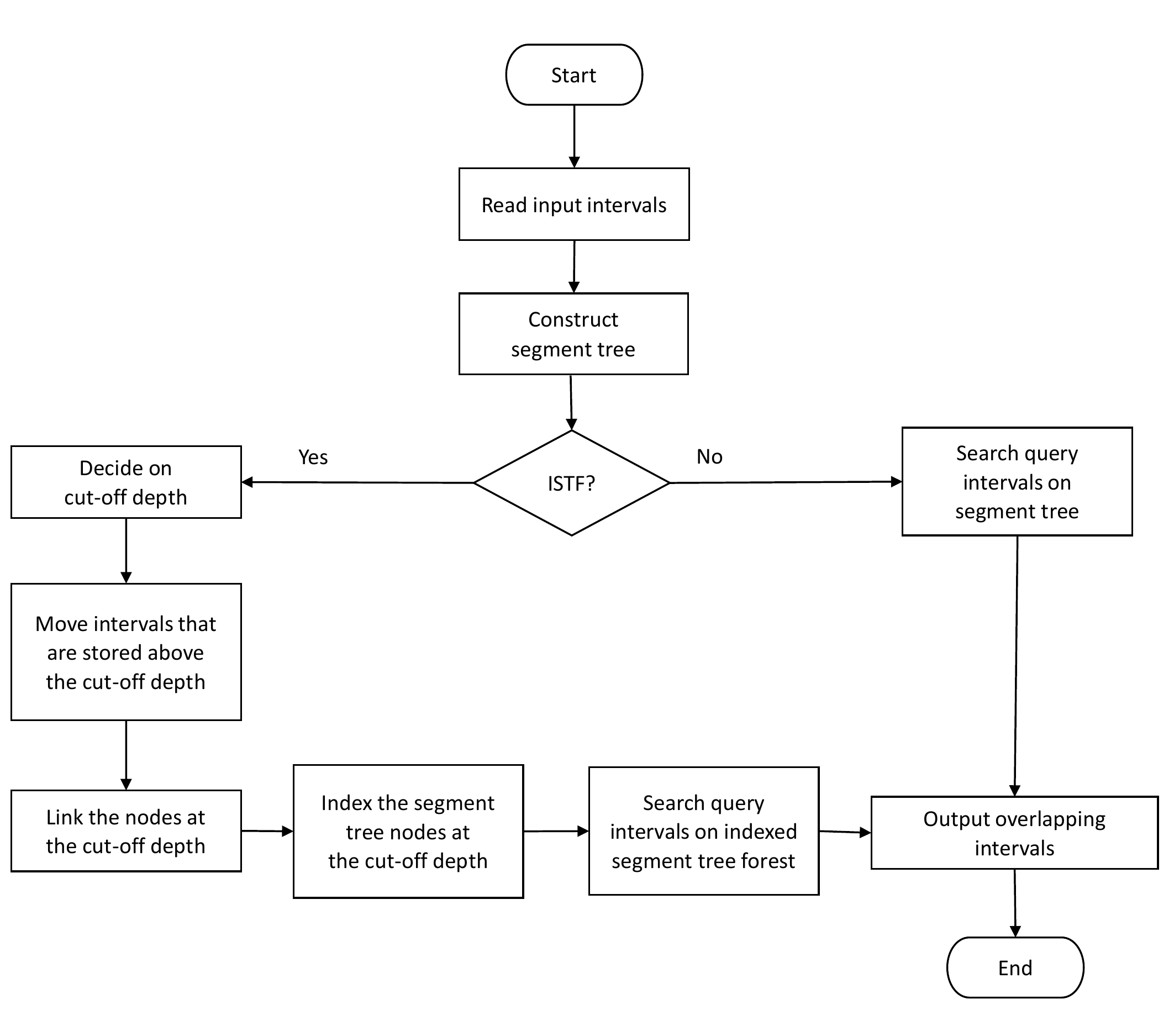}
\caption{Work-flow for finding jointly overlapping intervals of $n$ interval sets, displays the proposed data structures: segment tree and indexed segment tree forest, and the key steps on them.}
\label{Figure1}
\end{figure}

\begin{figure}
\centering
\includegraphics[width=0.48\textwidth]{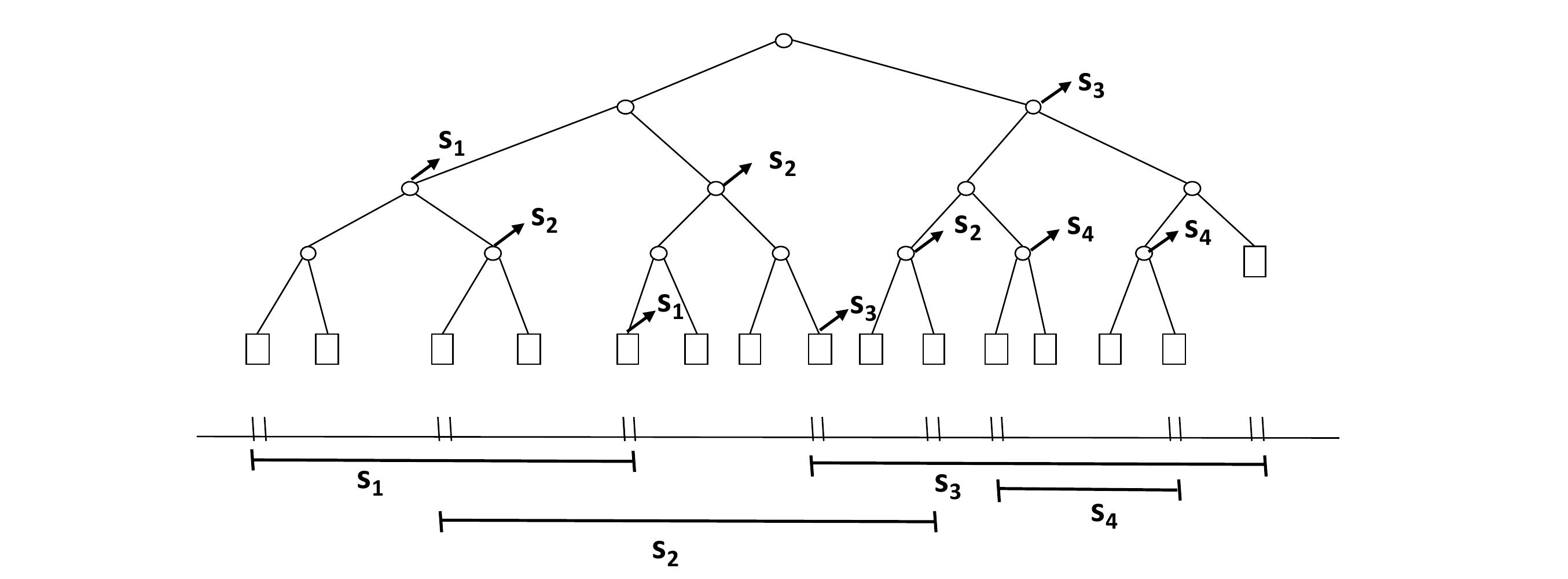}
\caption{Intervals ($s_{1},s_{2},s_{3},s_{4}$) are stored in the nodes. The arrows from the nodes point to their canonical subsets.}
\label{Figure2}
\end{figure}

\begin{figure}
\centering
\includegraphics[width=0.5\textwidth]{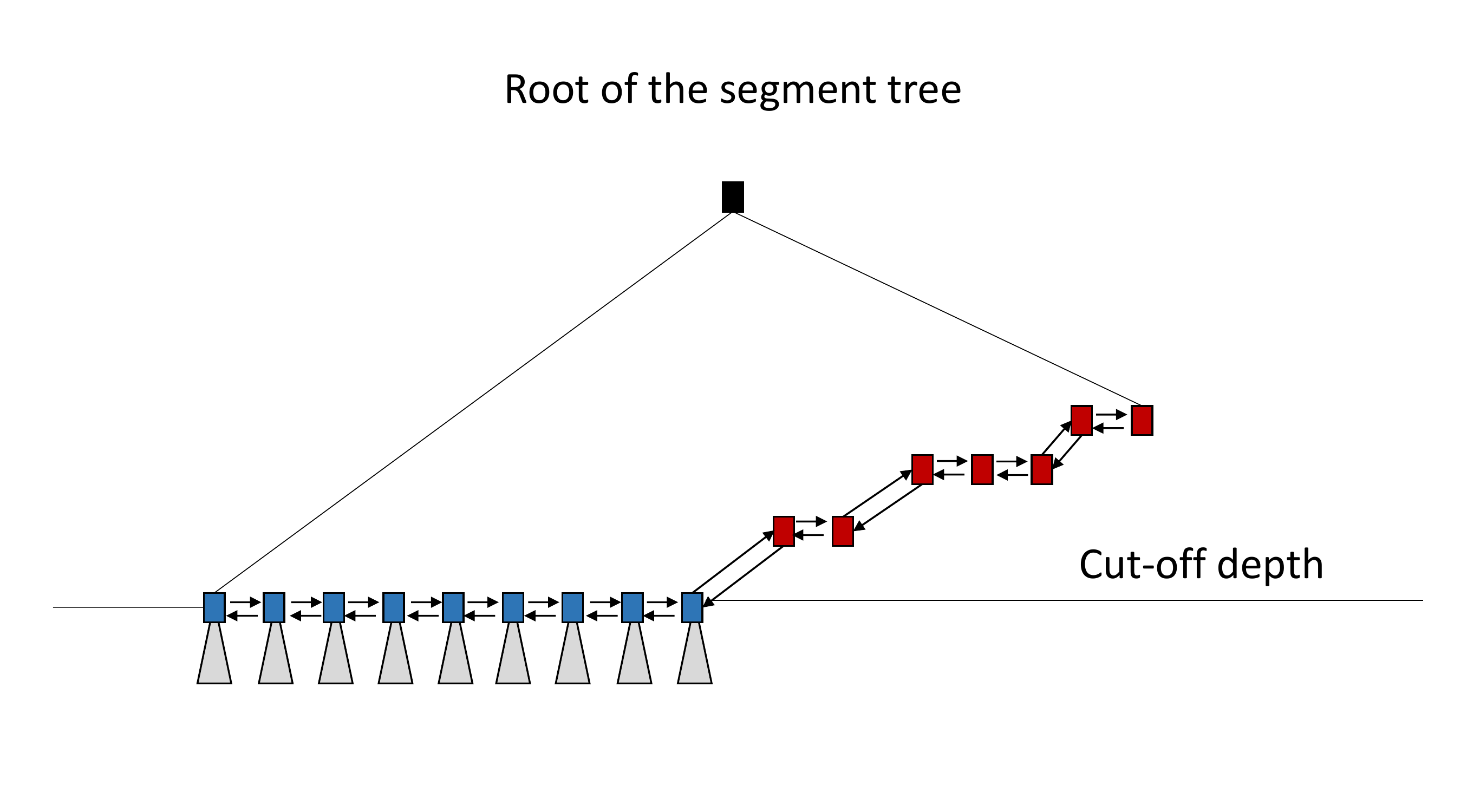}
\caption{Segment tree nodes (blue) at the cut-off depth and
segment tree nodes (red) with no children above the cut-off depth
are stored in our segment tree forest. To enhance fast access,
these stored segment tree nodes are connected to each other
through forward and backward links.}
\label{Figure3}
\end{figure}

\begin{figure}
\centering
\includegraphics[width=0.5\textwidth]{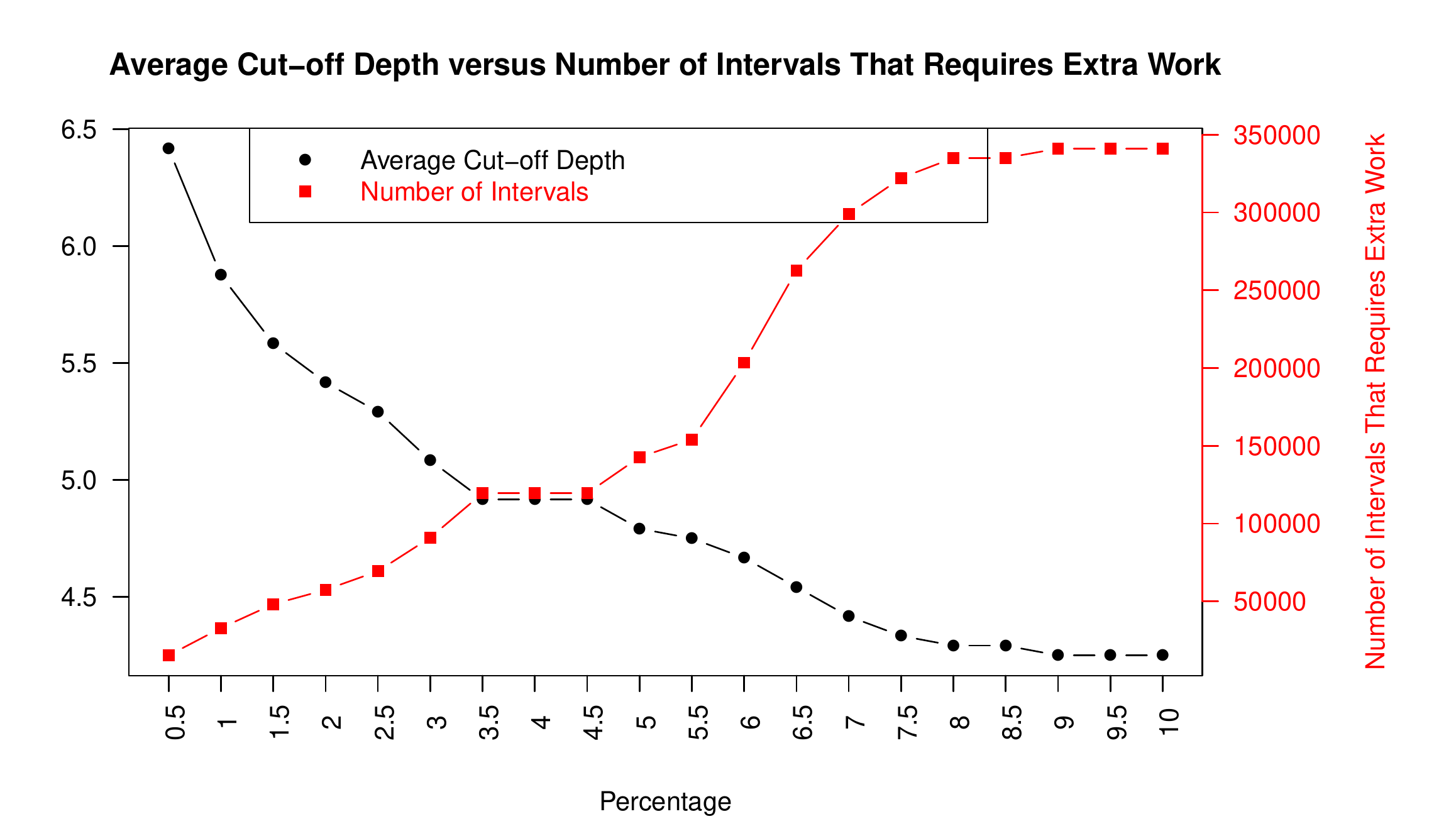}
\caption{We analyze the effect of percentage parameter on the average cut-off depth and the number of intervals that require extra work. This analysis shows that as we increase the percentage parameter, as expected, the number of intervals that need to be moved increases and average cut-off depth decreases.} 
\label{Figure4}
\end{figure}

\begin{figure}
\centering
\includegraphics[width=0.5\textwidth]{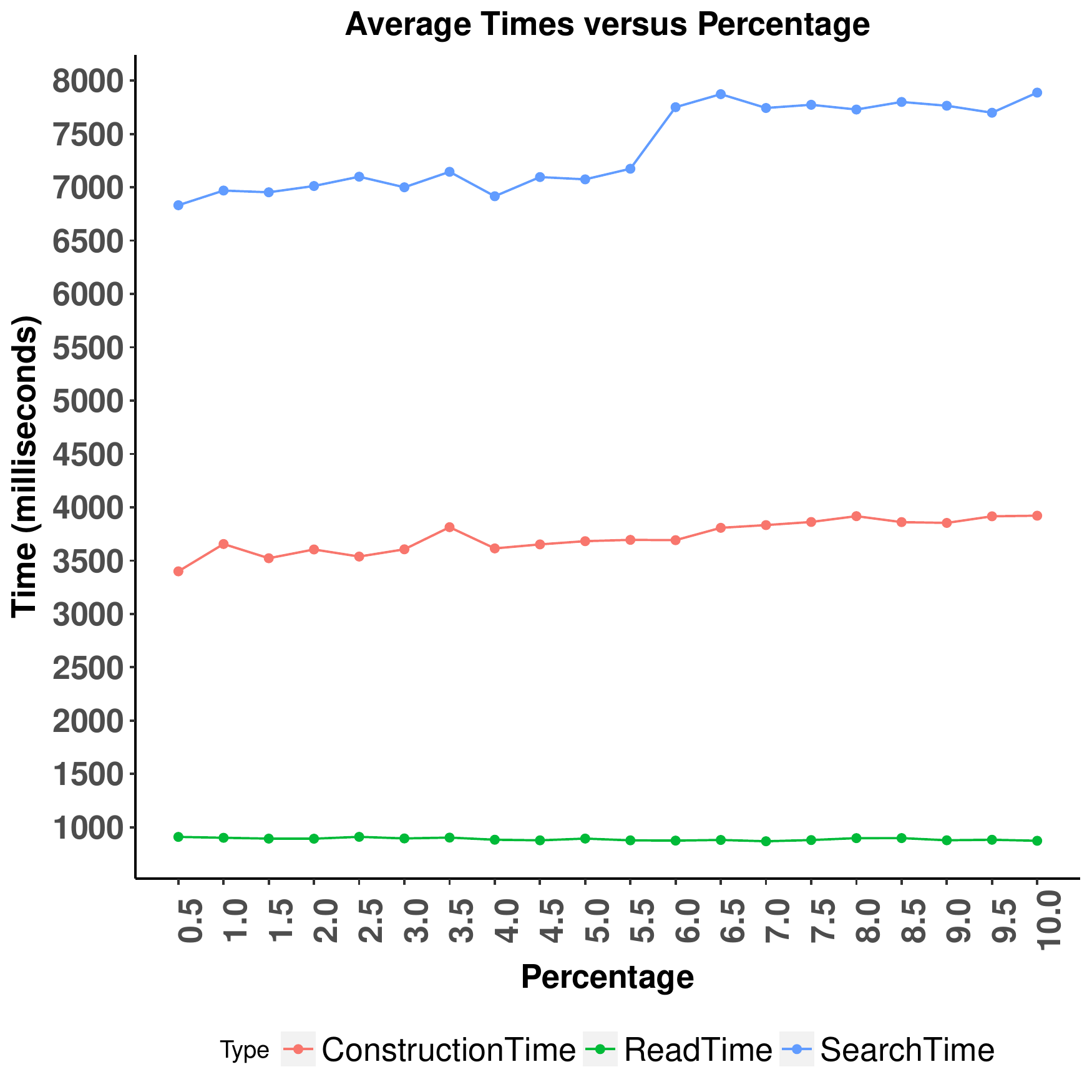}
\caption{We analyze the effect of percentage parameter on the average runtimes. These runtimes are averaged over $50$ runs using $2.3$ million intervals files. This analysis shows that as we increase the percentage parameter, as expected, average read time does not change but the construction time and query time are increased.} 
\label{Figure5}
\end{figure}


\begin{figure}
\centering
\includegraphics[width=0.5\textwidth]{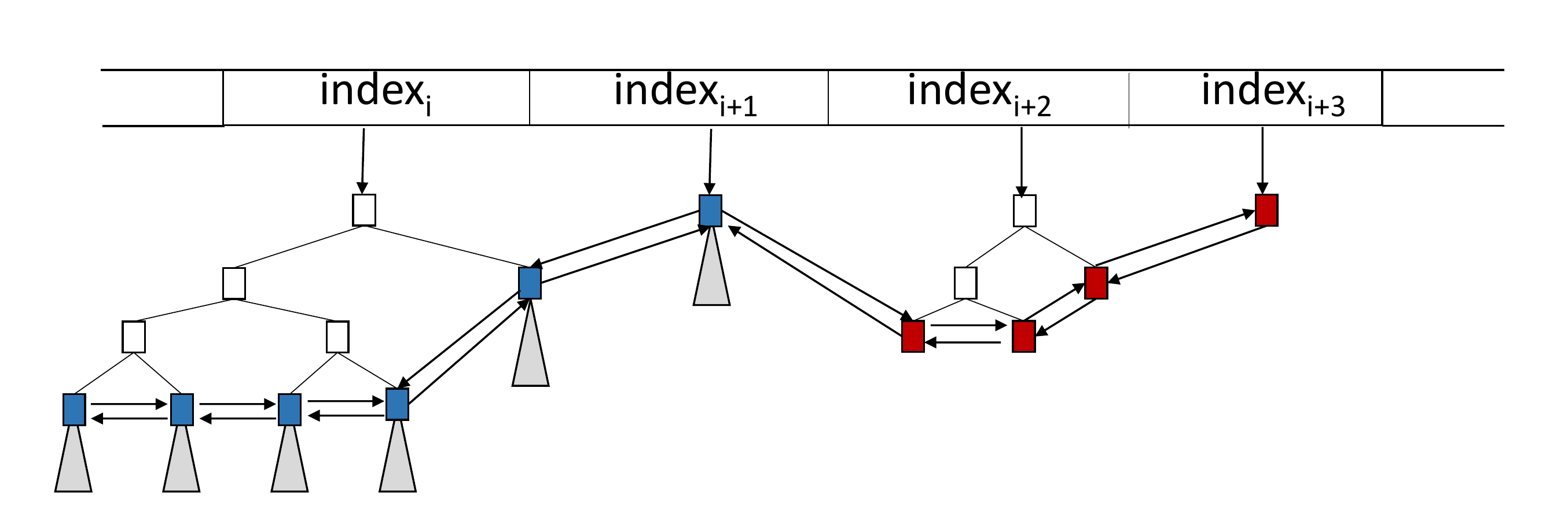}
\caption{Segment tree nodes with the same index are stored in a
BST and the root of BST is indexed. Blue and red colored
nodes are original segment tree nodes which are linked to each
other. Blue colored nodes are in fact the roots of the segment
trees below them. Red colored nodes do not have any children.
Parents of these blue and red colored nodes are the artificial
nodes, if any.} 
\label{Figure6}
\end{figure}

\begin{figure}
\centering
\includegraphics[width=0.50\textwidth]{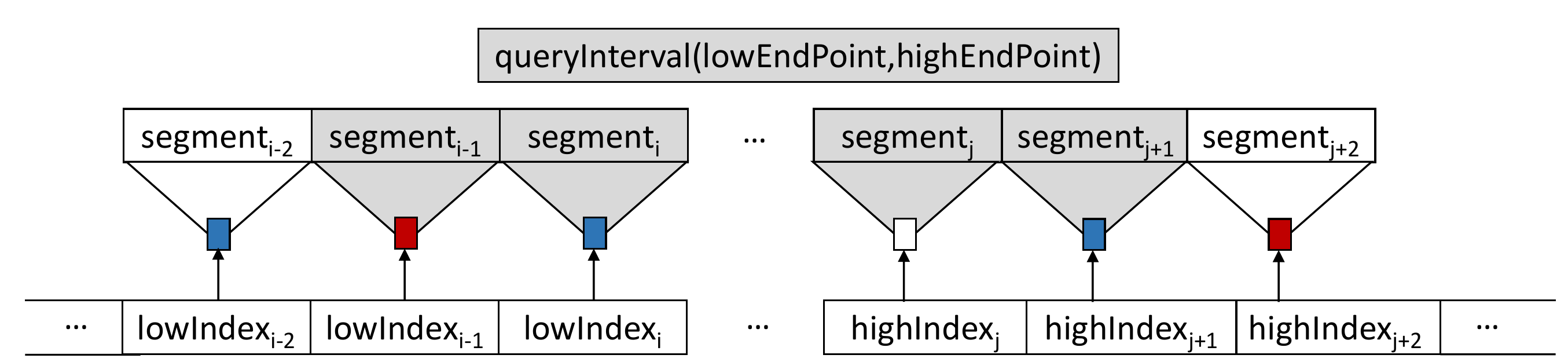}
\caption{Searching the nodes pointed by $lowIndex_{i}$ and
$highIndex_{j}$, the nodes in between them, and plus two more
nodes at most is enough.} 
\label{Figure7}
\end{figure}

\begin{figure}
\centering
\includegraphics[width=0.45\textwidth]{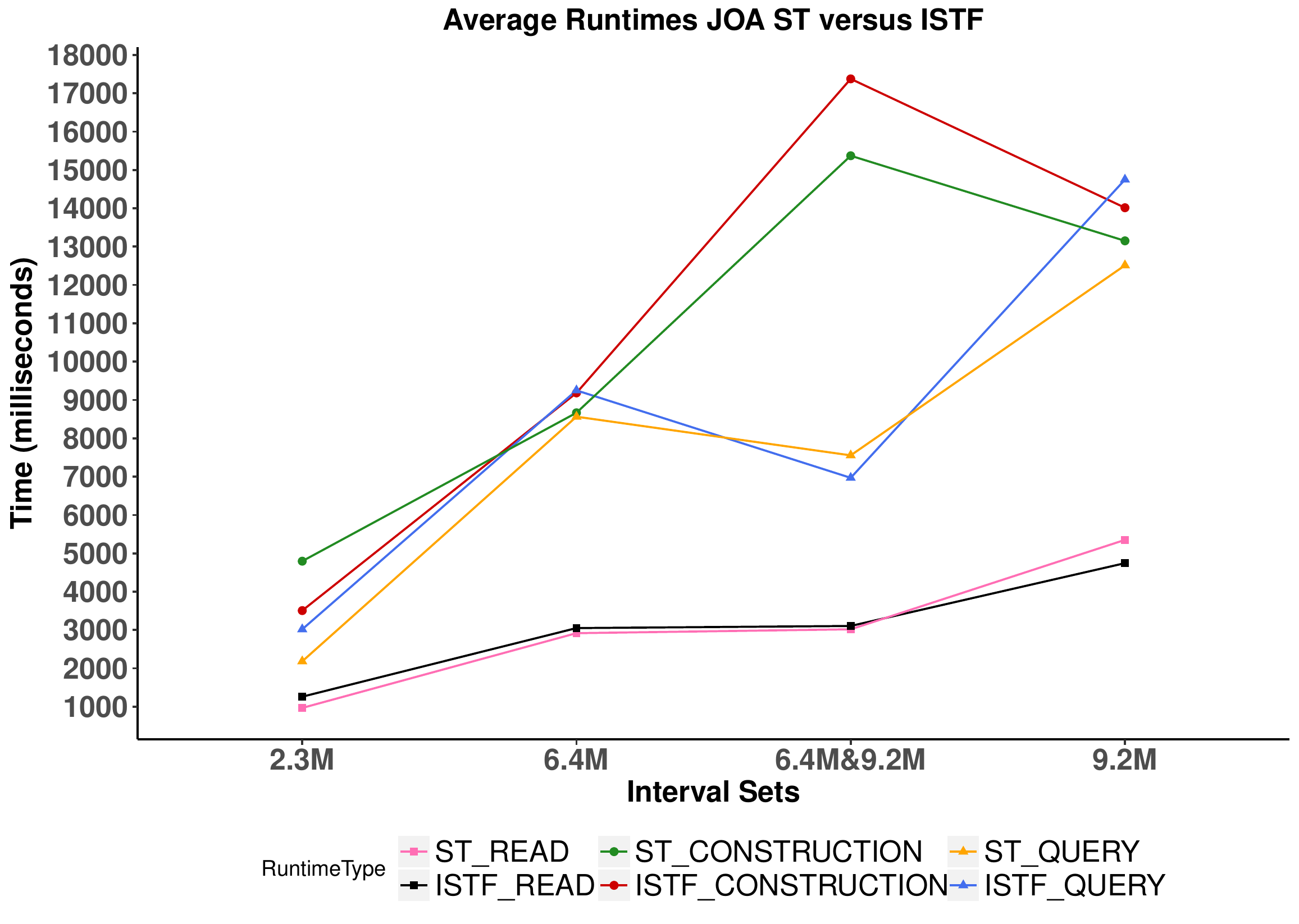}
\caption{We compare JOA segment tree and indexed segment tree forest w.r.t. read, construction and query time which are averaged over 50 runs. This comparison shows that segment tree and indexed segment tree forest are comparable to each other.} 
\label{Figure8}
\end{figure}

\begin{figure}
\centering
\includegraphics[width=0.40\textwidth]{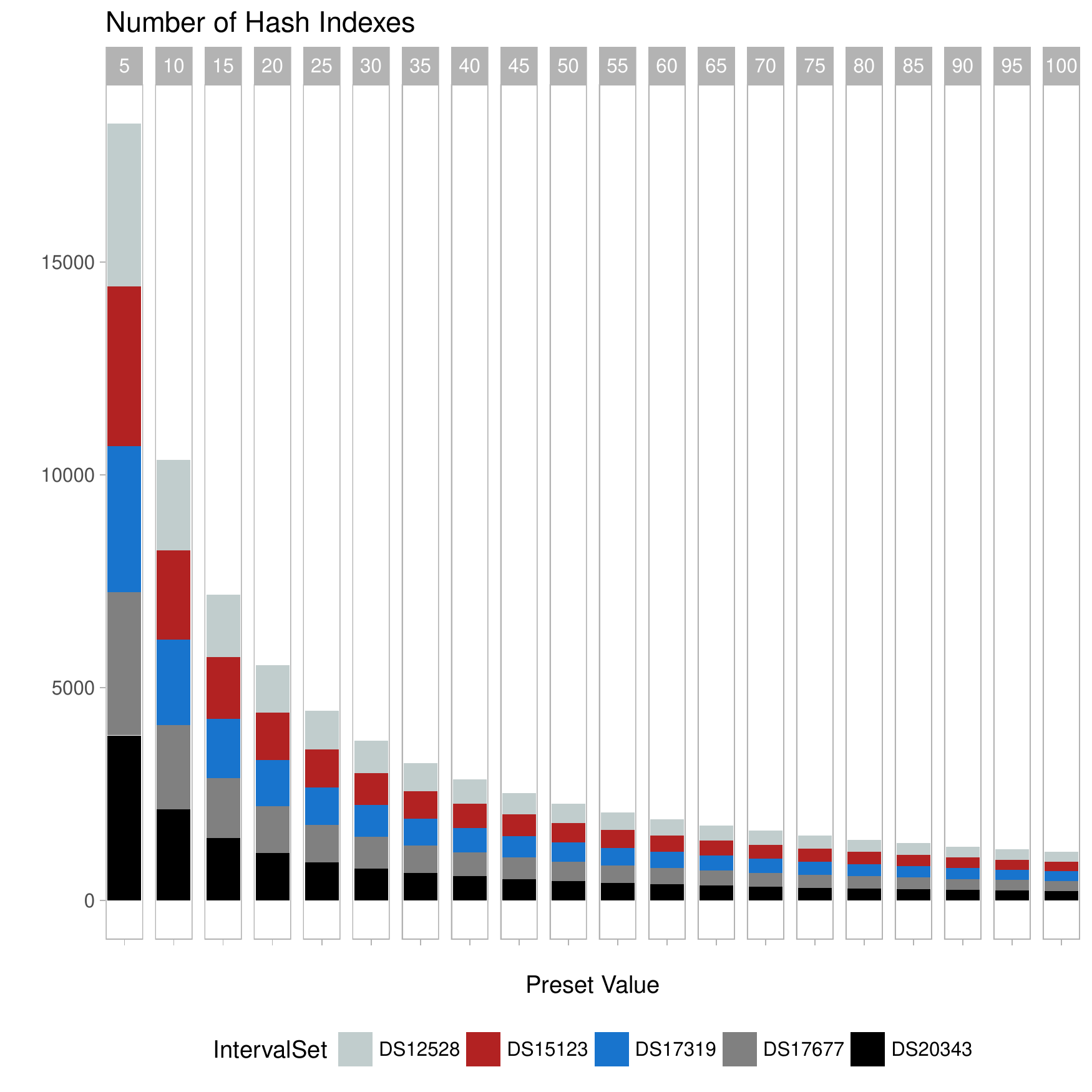}
\caption{Number of hash indexes for varying preset values. Preset values must be multiplied by 10K.} 
\label{Figure9}
\end{figure}

\begin{figure}
\centering
\includegraphics[width=0.4\textwidth]{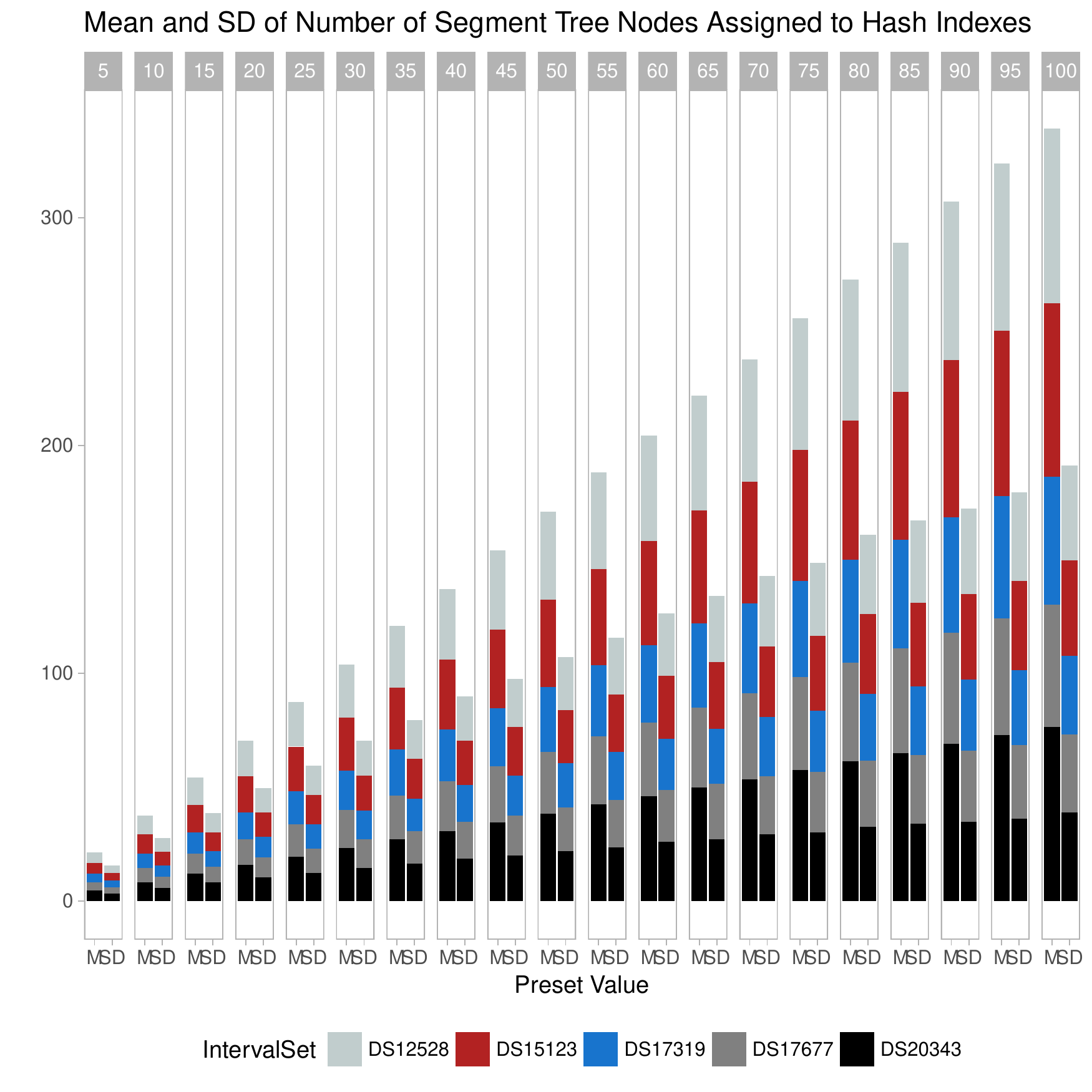}
\caption{Mean and standard deviation of number of segment  tree
nodes assigned to hash indexes for varying preset values. Preset
values must be multiplied by 10K.}
\label{Figure10}
\end{figure}




\end{backmatter}

\end{document}


\maketitle

\clearpage
\section{Indexed Segment Tree Forest Search Algorithms}
\FloatBarrier
\begin{algorithm}
\caption{search} \label{Algo:Search}
\begin{algorithmic}[1]
 \REQUIRE $queryIntervals$
 \REQUIRE $index2NodeMap$
 \REQUIRE $overlappingIntervalsList$
 \STATE $qOvIntList:  queryOverlappingIntervalsList$
 \FOR{each query interval}
    \STATE $qOvIntList \gets mainSearch(query,index2NodeMap,presetValue)$
    \STATE update $overlappingIntervalsList$ with $qOvIntList$
  \ENDFOR
\end{algorithmic}
\end{algorithm}

\begin{algorithm}
\caption{mainSearch} \label{Algo:MainSearch}
\begin{algorithmic}[1]
  \REQUIRE $query(lowEndPoint,highEndPoint)$
  \REQUIRE $index2NodeMap$
  \REQUIRE $presetValue > 0$
  \STATE $overlappingIntervals \gets \emptyset$
  \STATE $lowIndex \gets lowEndPoint/presetValue$
  \STATE $highIndex \gets highEndPoint/presetValue$
  \STATE $lowNode \gets index2NodeMap.get(lowIndex)$
  \IF{$lowNode \neq null$ and $linked(lowNode)$}
    \STATE $searchAtLinkedNode(lowNode,query,$ $overlappingIntervals)$
  \ELSIF{$lowNode \neq null$}
    \IF{$overlaps(query,lowNode)$}
        \STATE{$searchDownward(lowNode,query,$ $overlappingIntervals)$}
    \ENDIF
  \STATE $rightNode \gets findRightMostLnkd(lowNode)$
  \STATE $searchForward(rightNode.forwardNode,query,$ $overlappingIntervals)$
  \STATE $leftNode \gets findLeftMostLnkd(lowNode)$
  \STATE $searchBackward(leftNode.backwardNode,query,$ $overlappingIntervals)$
  \ELSE
    \STATE $lowerIndex \gets getLowerIndex(index2NodeMap,lowIndex)$
    \STATE $lowerNode = index2NodeMap.get(lowerIndex)$
    \IF{$lowerNode \neq null$}
        \STATE $searchAtLowerNode(lowerNode,query,$ $overlappingIntervals)$
    \ELSE
        \STATE $highNode \gets index2NodeMap.get(highIndex)$
        \IF{$highNode \neq null$ and $linked(highNode)$}
            \STATE $searchAtLinkedNode(highNode,query,$ $overlappingIntervals)$
        \ELSIF{$highNode \neq null$}
            \IF{$overlaps(query,highNode)$}
                \STATE{$searchDownward(highNode,query,$ $overlappingIntervals)$}
            \ENDIF  

            \STATE $rightNode \gets findRightMostLnkd(highNode)$
            \STATE $searchForward(rightNode.forwardNode,$ $query,overlappingIntervals)$
            \STATE $leftNode \gets findLeftMostLnkd(highNode)$
            \STATE $searchBackward(leftNode.backwardNode,$ $query,overlappingIntervals)$
        \ELSE
            \STATE $higherIndex \gets getHigherIndex(index2NodeMap,highIndex)$
            \STATE $higherNode = index2NodeMap.get(higherIndex)$
            \IF{$higherNode \neq null$}
                \STATE $searchAtHigherNode(higherNode,query,$  $overlappingIntervals)$
            \ENDIF
        \ENDIF
    \ENDIF
  \ENDIF
  \\
  \Return $overlappingIntervals$
\end{algorithmic}
\end{algorithm}

\begin{algorithm}
\caption{searchAtLinkedNode} \label{Algo:SearchAtLinkedNode}
\begin{algorithmic}[1]
  \REQUIRE $node$ is a linked original node
  \REQUIRE $query(lowEndPoint,highEndPoint)$
  \REQUIRE $overlappingIntervals$
  \STATE $searchForward(node,query,overlappingIntervals)$
  \STATE $searchBackward(node.backwardNode,query,$ $overlappingIntervals)$
\end{algorithmic}
\end{algorithm}

\begin{algorithm}
\caption{searchForward} \label{Algo:SearchForward}
\begin{algorithmic}[1]
  \REQUIRE $node$ is a linked original node
  \REQUIRE $query(lowEndPoint,highEndPoint)$
  \REQUIRE $overlappingIntervals$
  \STATE $low$: $lowEndPoint$
  \STATE $high$: $highEndPoint$
  \IF{$node  \neq null$ and $node.interval.low \leq high$}
    \IF {$low \leq node.interval.high$}
        \STATE add $node.canonicalSubset$ to $overlappingIntervals$
        \IF{$node.left \neq null$ and $low \leq node.left.interval.high$}
            \STATE $searchDownward(node.left,query,$ $overlappingIntervals)$
        \ENDIF
        \IF{$node.right \neq null$ and $node.right.interval.low \leq high$}
            \STATE $searchDownward(node.right,query,$ $overlappingIntervals)$
        \ENDIF
    \ENDIF
    \STATE $searchForward(node.forwardNode,query,$ $overlappingIntervals)$
  \ENDIF
\end{algorithmic}
\end{algorithm}

\begin{algorithm}
\caption{searchBackward} \label{Algo:SearchBackward}
\begin{algorithmic}[1]
  \REQUIRE $node$ is a linked original node
  \REQUIRE $query(lowEndPoint,highEndPoint)$
  \REQUIRE $overlappingIntervals$
  \STATE $low$: $lowEndPoint$
  \STATE $high$: $highEndPoint$
   \IF{$node \neq null$ and $low \leq node.interval.high$}
    \IF {$node.interval.low \leq high$}
        \STATE add $node.canonicalSubset$ to $overlappingIntervals$
        \IF{$node.left \neq null$ and $low \leq node.left.interval.high$}
            \STATE $searchDownward(node.left,query,$ $overlappingIntervals)$
        \ENDIF
        \IF{$node.right \neq null$ and $node.right.interval.low \leq high$}
            \STATE $searchDownward(node.right,query,$ $overlappingIntervals)$
        \ENDIF
    \ENDIF
    \STATE $searchBackward(node.backwardNode,query,$ $overlappingIntervals)$
  \ENDIF
\end{algorithmic}
\end{algorithm}

\begin{algorithm}
\caption{searchDownward} \label{Algo:SearchDownward}
\begin{algorithmic}[1]
  \REQUIRE $query(lowEndPoint,highEndPoint)$
  \REQUIRE $node \neq null$
  \REQUIRE $node$ and query overlaps
  \REQUIRE $overlappingIntervals$
  \STATE $low$: $lowEndPoint$
  \STATE $high$: $highEndPoint$
        \STATE Add $node.canonicalSubset$ to $overlappingIntervals$
        \IF{$node.left \neq null$ and $low \leq node.left.interval.high$}
            \STATE $searchDownward(node.left,query,$ $overlappingIntervals)$
        \ENDIF
        \IF{$node.right \neq null$ and $node.right.interval.low \leq high$}
            \STATE $searchDownward(node.right,query,$ $overlappingIntervals)$
        \ENDIF
\end{algorithmic}
\end{algorithm}

\begin{algorithm}
\caption{searchAtLowerNode} \label{Algo:SearchAtLowerNode}
\begin{algorithmic}[1]
  \REQUIRE $lowerNode \neq null$
  \REQUIRE $query(lowEndPoint,highEndPoint)$
  \REQUIRE $overlappingIntervals$
  \IF{$linked(lowerNode)$}
    \STATE  $searchForward(lowerNode,query,$ $overlappingIntervals)$
  \ELSE
    \IF{$overlaps(query,lowerNode)$}
        \STATE $searchDownward(lowerNode,query,$ $overlappingIntervals)$
    \ENDIF
    \STATE $node \gets findRightMostNode(lowerNode)$
    \STATE $searchForward(node.forwardNode,query,$ $overlappingIntervals)$
  \ENDIF
\end{algorithmic}
\end{algorithm}

\begin{algorithm}
\caption{searchAtHigherNode} \label{Algo:SearchAtHigherNode}
\begin{algorithmic}[1]
  \REQUIRE $higherNode \neq null$
  \REQUIRE $query(lowEndPoint,highEndPoint)$
  \REQUIRE $overlappingIntervals$
  \IF{$linked(higherNode)$}
    \STATE  $searchBackward(higherNode,query,$ $overlappingIntervals)$
  \ELSE
    \IF{$overlaps(query,higherNode)$}
        \STATE $searchDownward(higherNode,query,$ $overlappingIntervals)$
    \ENDIF
    \STATE $node \gets findLeftMostNode(higherNode)$
    \STATE $searchBackward(node.backwardNode,query,$ $overlappingIntervals)$
  \ENDIF
\end{algorithmic}
\end{algorithm}
\FloatBarrier

\section{Jointly Overlapping Intervals for 141 ENCODE Dnase Hypersensitive Sites}
{\bf Supplementary Table 1}: An additional case study for JOA using 141 ENCODE Dnase hypersensitive sites. We found and supplied all
jointly overlapping intervals for these 141 Dnase hypesensitive sites files in. Excel file: SupplementaryTable1.xlsx\\

\section{JOA Memory Usage}
\begin{table*}[!htbp]
\begin{center}
\fontsize{10}{11}\selectfont
\begin{tabular}
{| c | r | r |} \hline
\multicolumn{1}{|c|}{\multirow{3}{*}{\parbox{3.5cm}{\centering Simulations\\$1^{st}$ Scenario\\(\#ofFiles,\#ofIntervals)}}}& \multicolumn{2}{c|}{Memory Usage in MBs}    \\  \cline{2-3}
\multicolumn{1}{|c|}{} & \multicolumn{1}{c|}{\multirow{2}{*}{JOA ST}} & \multicolumn{1}{c|}{\multirow{2}{*}{JOA ISTF}}   \\ 
\multicolumn{1}{|c|}{} & \multicolumn{1}{c|}{} &  \multicolumn{1}{c|}{}  \\ \cline{1-3}
(2,100000)	& 149 	& 160	\\ \cline{1-3}
(4,100000) 	& 297 	& 321	\\ \cline{1-3}
(8,100000)	& 97  	& 117	\\ \cline{1-3}
(16,100000) 	& 84 	& 115 	\\ \cline{1-3}
(32,100000) 	& 80 	& 236	\\ \cline{1-3}
(64,100000) 	& 551 	& 401	\\ \cline{1-3}
(128,100000) & 518 	& 268	\\ \cline{1-3}
(256,100000) & 380 	& 135	\\ \cline{1-3}
(512,100000) & 604 	& 594 	\\ \cline{1-3}
\end{tabular}
\end{center}
Supplementary Table 2. Memory Usage of JOA in MBs for the $1^{st}$ scenario simulated datasets. \label{MemoryUsage_JOA_Simulations_Scenario1_Table}
\end{table*}

\begin{table*}[!htbp]
\begin{center}
\fontsize{10}{11}\selectfont
\begin{tabular}
{| c | r | r |} \hline
\multicolumn{1}{|c|}{\multirow{3}{*}{\parbox{3.5cm}{\centering Simulations\\$2^{nd}$ Scenario\\(\#ofFiles,\#ofIntervals)}}}& \multicolumn{2}{c|}{Memory Usage in MBs}    \\  \cline{2-3}
\multicolumn{1}{|c|}{} & \multicolumn{1}{c|}{\multirow{2}{*}{JOA ST}} & \multicolumn{1}{c|}{\multirow{2}{*}{JOA ISTF}}   \\ 
\multicolumn{1}{|c|}{} & \multicolumn{1}{c|}{} &  \multicolumn{1}{c|}{}  \\ \cline{1-3}
(2,1M)	& 935 	& 975	\\ \cline{1-3}
(2,2M) 	& 1894 	& 1944	\\ \cline{1-3}
(2,4M)	& 4171  	& 4210	\\ \cline{1-3}
(2,8M) 	& 9039 	& 9179 	\\ \cline{1-3}
(2,16M) 	& 21343 	& 18659	\\ \cline{1-3}
\end{tabular}
\end{center}
Supplementary Table 3. Memory Usage of JOA in MBs for the $2^{nd}$ scenario simulated datasets. \label{MemoryUsage_JOA_Simulations_Scenario2_Table}
\end{table*}